\documentclass[aps,prl,preprint,superscriptaddress]{revtex4-1}
               
\usepackage{graphicx}   
\usepackage{dcolumn}    
\usepackage{bm}         
\usepackage{amsmath}
\usepackage{amssymb}
\usepackage[usenames]{color}      

\begin{document}

\title{Thermohaline interleaving induced by horizontal temperature and salinity gradients from above}

\author{Junyi Li}
\affiliation{State Key Laboratory for Turbulence and Complex Systems and Department of Mechanics and Engineering Science, Beijing Innovation Center for Engineering Science and Advanced Technology, College of Engineering, and Institute of Ocean Research, Peking University, Beijing 100871, China}

\author{Yantao Yang}\email{yantao.yang@pku.edu.cn}
\affiliation{State Key Laboratory for Turbulence and Complex Systems and Department of Mechanics and Engineering Science, Beijing Innovation Center for Engineering Science and Advanced Technology, College of Engineering, and Institute of Ocean Research, Peking University, Beijing 100871, China}
       
\date{\today}
	
\begin{abstract}

In the Ocean, thermohaline intrusions and interleaving layers occur within the water mass fronts with horizontal temperature and salinity gradients, which provide an important horizontal mixing mechanism. Here we report a new type of thermohaline intrusion which is driven by the horizontal temperature and salinity gradients in the fluid layers at adjacent depths due to the different double diffusive mixing rates in the vertical direction. Once established, the intrusion layers share similar behaviors as those found within the gradient regions. Such intrusion process generates extra horizontal heat and salinity fluxes towards the cold and fresh side, but transfer density anomaly towards the warm and salty side. These findings greatly extend the circumstance where thermohaline intrusions may be observed. 

\end{abstract}

\maketitle

Water mass fronts with lateral gradients of temperature and salinity are omnipresent in the Ocean, and thermohaline intrusions and interleaving often occur in these areas~\cite{woods1986,holbrook2003,ruddickrichards2003,timmermans2020}. Such processes generate fine-scale thermohaline structures, drive lateral flux and mixing, and affect the evolution of mesoscale vortices~\cite{merryfield2000,lee2004,ruddick2010,sarkar2015,bebieva2019,tang2019}. Thus, numerous studies have been carried out to understand the formation mechanism and transport properties of thermohaline intrusions and interleaving~\cite{ruddickkerr2003,ruddick2003}, and double diffusive mixing is believed to be one of the key mechanisms~\cite{radko2013,radko2017}.

Double diffusive convection (DDC) happens when fluid density depends on two scalars which have very different molecular diffusivities and experience certain gradients. The difference in diffusivities of temperature and salinity is usually two orders of magnitude, and the Ocean is prone to double diffusive mixing~\cite{you2002}. Since the groundbreaking work of Stern~\cite{stern1967}, various studies have proved that double diffusive mixing can generate interleaving unstable modes for different frontal configurations, which agree with the field observations in many aspects~\cite{ruddickkerr2003,ruddick2003,radko2013}. 

In most of the existing studies, the interleaving and layering happen within the water mass which experiences the vertical and horizontal gradients of temperature and salinity, e.g. see~\cite{ruddick2003,krishnamurti2006,simeonovstern2007,hebert2011}. Here by direct numerical simulation (DNS), we reveal that thermohaline interleaving can also develop outside the water mass with horizontal gradients. Specifically, we will show that horizontal gradients of temperature and salinity are sufficient to induce thermohaline interleaving and spontaneous layering underneath, and the double diffusive mixing plays a critical role in such process. 

We consider a Cartesian box with the height $H$ in the $z$-direction and a length $L$ in the $y$-direction. The gravity is in the negative $z$-direction. The flow is statistically homogeneous in the $x$-direction. At the top boundary both temperature and salinity increase linearly along the $y$-direction, which simultaneously drive the convection flow in the domain. Hereafter, we respectively refer to the $(x,\,y,\,z)$ directions as the spanwise, streamwise and vertical directions, since the convection motions are mainly in the $(y,\,z)$-plane, as shown later.

The Oberbeck-Boussinesq approximation is employed. That is, the fluid density depends linearly on temperature and salinity as $\rho=\rho_0\left[ 1-\beta_T (T-T_0) + \beta_S (S-S_0) \right]$. Here $\rho$ is density, $T$ is temperature, $S$ is salinity, $\beta_T$ is the thermal expansion coefficient, and $\beta_S$ is the contraction coefficient of salinity, respectively. The subscript ``0'' indicates the value at a reference state. Furthermore, the governing equations read
\begin{eqnarray}
	\partial_t \mathbf{u} + \mathbf{u} \cdot \nabla\mathbf{u} 
	    &=&  - \frac{1}{\rho}\nabla p + \nu  \nabla^2 \mathbf{u} +g(\beta_T T-\beta_S S)\mathbf{e}_z, \label{eq:u}\\
	\partial_t T + \mathbf{u} \cdot \nabla T &=& \kappa_T \nabla^2 T, \label{eq:t} \\
	\partial_t S + \mathbf{u} \cdot \nabla S &=& \kappa_S \nabla^2 S, \label{eq:s}
\end{eqnarray}
where $\mathbf{u}$ is velocity, $p$ is pressure, $\nu$ is kinematic viscosity, $g$ is the gravitational acceleration, $\mathbf{e}_z$ is the unit vector in the $z$-direction, and $\kappa_T$ and $\kappa_S$ are the two molecular diffusivities, respectively. The continuity equation for incompressible flow is $\nabla\cdot\mathbf{u}=0$.

The flow quantities are nondimensionalized by the free-fall velocity $\sqrt{g\beta_S \Delta_S H}$, the domain height $H$, and the total scalar increments $\Delta_T$ and $\Delta_S$ along the top surface. The Prandtl number $Pr=\nu/\kappa_T$ is fixed at $7$, and the Schmidt number $Sc=\nu/\kappa_S$ at $21$, respectively. Note that $Sc=21$ is much smaller than the typical value in the Ocean, saying $700\sim1000$. However, large $Sc$ requires very fine grids and presents a big challenge for DNS. Therefore, a smaller $Sc=21$ is chosen in the current study, which is a common treatment in DDC simulations~\cite{stellmach2011,paparella2012}. The thermal Rayleigh number is defined as $Ra = (g\beta_T H^3 \Delta_T)/(\nu\kappa_T)$. The relative strength of the salinity gradient compared to that of temperature gradient can be measured by the density ratio $\Lambda = (\beta_T \Delta_T) / (\beta_S \Delta_S)$, i.e. the ratio of the density anomaly induced by the salinity difference to that by the temperature difference. Here we set $\Lambda=1$ so that the effects of temperature and salinity on the density compensate each other. The aspect ratio of the domain $\Gamma=L/H$ is fixed at 4. 

The governing equations~\eqref{eq:u}-\eqref{eq:s} are numerically solved by our in-house code with the finite-difference and fraction of time step method, which has been extensively used for wall-turbulence and convection flows~\citep{multigrid2015}. The bottom wall and two end walls in the streamwise direction are no-slip for velocity and adiabatic for two scalars. The top surface is free-slip and has uniform gradients for $T$ and $S$ in the $y$-direction. In the spanwise direction we fix the width as $H$ and use the periodic boundary conditions. Initially, the fluid is at rest and with uniform temperature and salinity distributions at $\Delta T/2$ and $\Delta S/2$, respectively. Small perturbations are added to both scalar fields initially to trigger the flow. Three cases were run with increasing $Ra=10^8$, $10^9$, and $10^{10}$, respectively.

\begin{figure}
	\centering
	\includegraphics[width=0.5\textwidth]{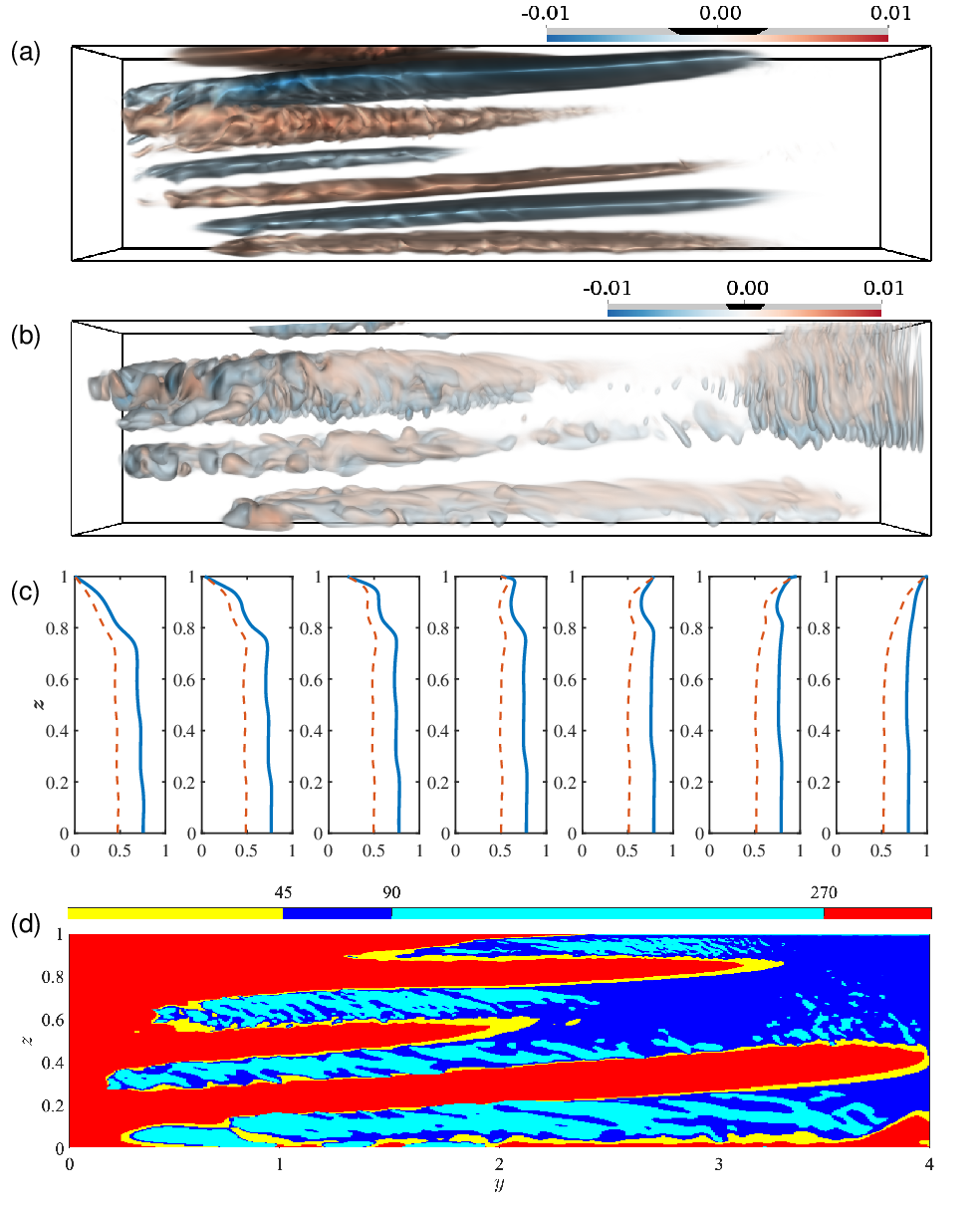}%
	\caption{Thermohaline interleaving for the case with $Ra=10^9$ depicted by the volume renderings of (a) the streamwise velocity and (b) the vertical velocity, in which the positive $y$-direction is from left to right. Panels (c) and (d) display some mean statistics of  the flow field shown in (a) and (b). (c) The spanwise-averaged scalar profiles at seven streamwise locations, i.e. $y=iL/8$, with $i=1...7$ from left to right. Temperature is shown by red dashed lines and salinity by blue solid lines, respectively. (d) The Turner angle $Tu$ computed by the spanwise-averaged scalar gradients.} 
	\label{fig1}
\end{figure}

Due to the opposite effects on density of two scalars, the temperature and salinity gradients over the top surface drive the convection flows in the different directions. Specifically, the cold fluid descends in one end while the salty fluid in the other end. Considering the difference in diffusivities, one may expect that the convection motions driven by two scalars extend to different depths below the top surface. The simulations reveal that, however, the flow field is much more complex. In Figs.~\ref{fig1}a and \ref{fig1}b we show the typical flow morphology by the three-dimensional volume rendering of two velocity components for the case with $Ra=10^9$. A stack of distinct layers develop in the domain, and the layering is stronger at the left side, saying with smaller surface temperature and salinity. For this case seven layers are visible. The layers extend upward towards the positive streamwise direction and have alternating flow directions with a positive or negative streamwise velocity, which is a clear indication of interleaving. Fig.~\ref{fig1}b further reveals that, vigorous vertical motions only happen at certain regions in the flow domain and exhibit more small-scale structures compared to the streamwise velocity.

Actually, fingering double diffusive convection is one of the main mechanisms which drive the vertical motions depicted in Fig.~\ref{fig1}b. To demonstrate this, we plot the mean scalar profiles at seven different streamwise locations in Fig.~\ref{fig1}c, which are calculated by taking the spanwise average of the flow field shown in Figs.~\ref{fig1}a and \ref{fig1}b. Within the upper half of the domain, different stratification patterns develop at two ends. On the left, both scalars increases with depth, favoring the diffusive type of DDC. While on the other end, the two scalars decrease with depth and salt fingers emerge, e.g.~see the vertically oriented thin finger-like structures at the upper-right corner in Fig.~\ref{fig1}b. The slabs shown in Fig.~\ref{fig1}b also consist of finger structures which are driven by the local stratification imposed by the oppositely moving interleaving layers. Specifically, the leftward moving layers carry higher temperature and salinity than the rightward moving layers below, and fingering convection occurs in between, which corresponds to the vertical motions in the left part of Fig.~\ref{fig1}b. This finger convection between interleaving layers is another characteristic phenomenon of thermohaline intrusion~\cite{ruddick2003,radko2013}.

The different types of DDC are better indicated by the Turner angle, which is defined by the vertical gradients of the mean scalars as $Tu=135^\circ-arg(\beta_S \partial_z \overline{S} + i\beta_T \partial_z \overline{T})$~\cite{ruddick1983}. Fingering DDC usually happens in the range $45^\circ<Tu<90^\circ$, while diffusive DDC in the range $270^\circ<Tu<315^\circ$, respectively. In Fig.~\ref{fig1}d we show $Tu$ for the same flow field, where the mean gradients are calculated after spanwise averaging. The region with $45^\circ<Tu<90^\circ$ agrees with the vertical motions in Fig.~\ref{fig1}b, namely, those vertical motions indeed are fingering convection. At the upper-left corner diffusive DDC dominates. The layers with $270^\circ<Tu<315^\circ$ correspond to the interfaces above the leftward moving warm and salty currents and below the rightward moving cold and fresh ones.

\begin{figure}
	\centering
	\includegraphics[width=0.5\textwidth]{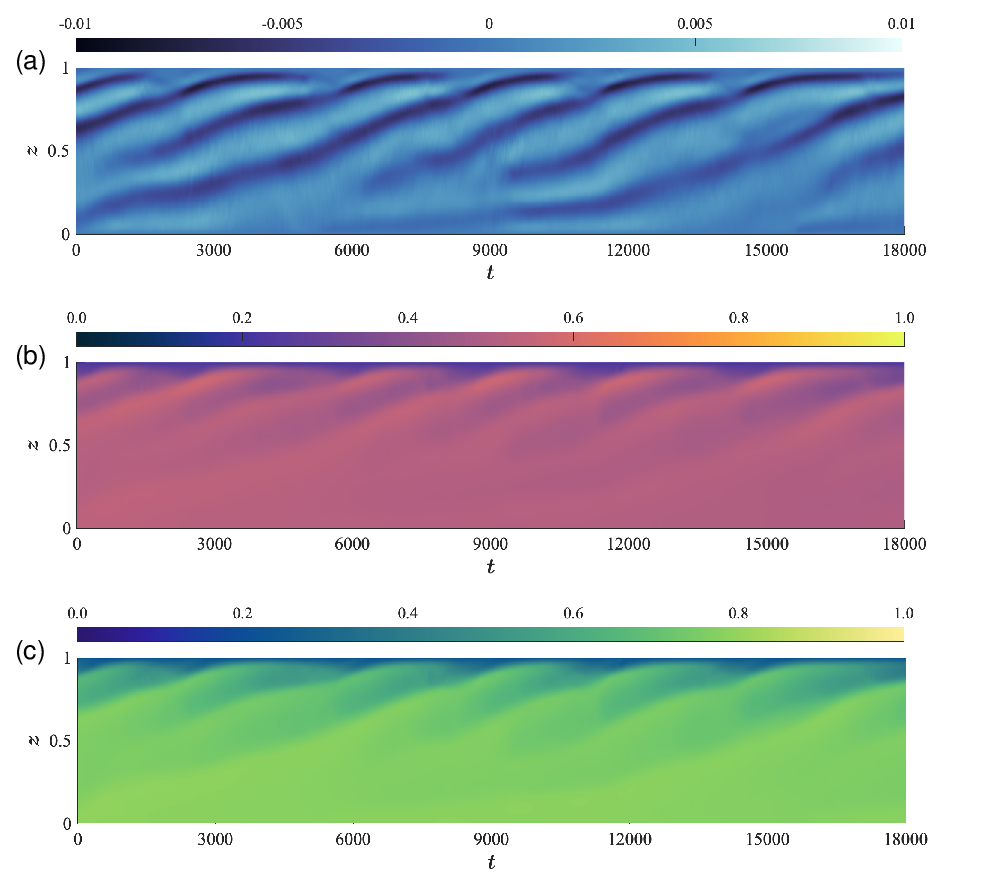}%
	\caption{The upward shifting of the interleaving layers depicted by the time evolution of the spanwise-averaged profiles of (a) the streamwise velocity, (b) temperature, and (c) salinity at the streamwise location $y=3L/8$ for the case with $Ra=10^9$, respectively.} 
	\label{fig2}
\end{figure}

The interleaving currents are nearly horizontal, with a small inclined angle about $3^\circ$. Moreover, our numerical results reveal that the vertical location of each layer is not constant. Rather, those layers shift upward over a very large time-scale, which is clearly shown in Fig.~\ref{fig2} by the time evolution of the vertical profiles of three flow quantities at the streamwise location $y=3/8L$, i.e. where the interleaving currents are strongest. It takes about $10000$ non-dimensional time units for one layer to move from bottom to the top surface. Detailed investigation reveals that close to the bottom boundary the total density difference between the two ends in the streamwise direction periodically changes sign, and new layers are generated accordingly with alternating moving directions. 

\begin{figure}
	\centering
	\includegraphics[width=0.48\textwidth]{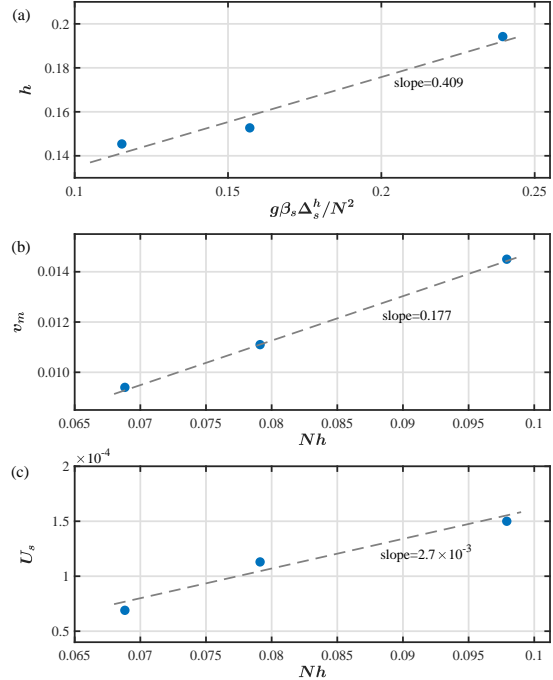}%
	\caption{(a) Layer thickness $h$ versus $(g\beta_S\Delta^h_S)/N^2$. (b, c) The maximal current velocity $v_m$ and the upward shifting velocity of layers $U_s$ versus $Nh$. } 
	\label{fig3}
\end{figure}

Key parameters of interests for intrusion are the thickness and strength or the current velocity of the interleaving layers. We use the spanwise averaged profiles at $y=3L/8$ to calculate the layer thickness $h$ by averaging over time and different layers, and the current strength by the maximal horizontal velocity $v_m$ of all layers. Despite the different flow configuration, the layer thickness $h$ in the current study agrees with the model proposed by Ruddick and Turner~\cite{ruddickturner1979}, where the authors use an energy analysis and arrive the scaling law of $h \sim C_h(g\beta_S\Delta^h_S)/N^2$. Here, $C_h$ is some constant, $\Delta^h_S$ is the lateral variation of salinity within the water mass, and $N=\sqrt{-(g/\rho)\partial_z\rho }$ is the buoyancy frequency, respectively. In Ruddick and Turner's experiments~\cite{ruddickturner1979}, both the lateral salinity variation and the vertical density stratification are carefully set as the control parameters, which is not the case in our simulations. Instead, we take $\Delta^h_S$ as the averaged salinity difference between two end walls in the streamwise direction, and calculate $N$ by the mean density difference between the top and bottom boundaries. In Fig.~\ref{fig3}a we plot $h$ versus $(g\beta_S\Delta^h_S)/N^2$, and indeed the data follow the scaling law with $C_h\approx0.409$. As shown in Fig.~\ref{fig3}b, we find that the behavior of the maximal current velocity $v_m$ is consistent with the model developed for tilted DNS simulations~\cite{simeonovstern2007},  saying $v_m\sim C_v N h$ with $C_v\approx0.177$, see Fig.~\ref{fig3}b.

The upward shifting of the intrusion layers over time, as shown in Fig.~\ref{fig2}, is induced by the continuous generation of new layers near the bottom of the domain. As new layers emerge, existing layers are pushed upward. Then one expects that the upward shifting velocity of the layers $U_s$ should exhibit similar scaling as the current velocity. Here $U_s$ can be easily measured from the contours in Fig.~\ref{fig2}. In Fig.~\ref{fig3}c we plot $U_s$ versus $Nh$ as done for $v_m$ in Fig.~\ref{fig3}b, and indeed $U_s$ also follows a roughly linear dependence on $Nh$. Moreover, the shifting velocity $U_s$ is much smaller than the current velocity $v_m$. That is, the layer shifting happens on the time scale much larger than that of the flow motion within the intrusion currents. 
 
\begin{figure}
	\centering
	\includegraphics[width=0.45\textwidth]{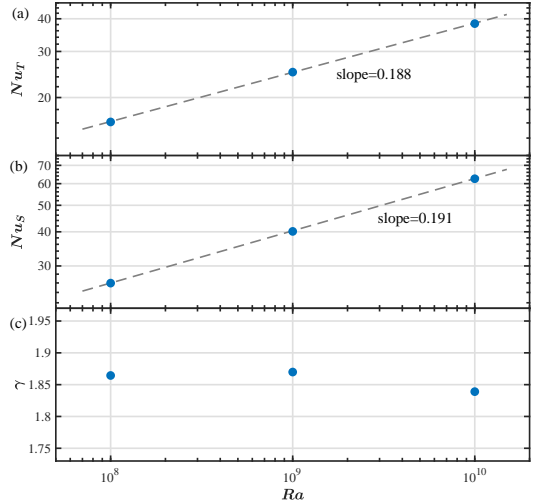}%
	\caption{The global horizontal fluxes of (a) temperature and (b) salinity versus the Rayleigh number $Ra$. Here the fluxes are measured by the corresponding Nusselt number $Nu$. (c) The horizontal density flux ratio $\gamma$ versus $Ra$.} 
	\label{fig4}
\end{figure}

Finally, we proceed to analyze the global transport properties in the horizontal direction. Due to our specific flow configuration, both heat and salinity are transported into the domain from the right half of the top surface with higher temperature and salinity, then horizontally towards the left side within the domain bulk, and finally out from the domain over the left half of the top surface with lower temperature and salinity. Thus, the global transfer of the two components is horizontal and in the opposite direction of the surface scalar gradients. We measure these global fluxes by Nusselt number defined as $Nu_\zeta = \left|\langle v\zeta \rangle - \kappa_\zeta \partial_y \langle \zeta \rangle\right|/(\kappa_\zeta\Delta_\zeta{L}^{-1})$ with $\zeta=T$ or $S$, in which $\langle\rangle$ denotes the average over the mid-plane $y=L/2$ and time. The dependences of $Nu_T$ and $Nu_S$ on $Ra$ are plotted in Figs.~\ref{fig4}a and \ref{fig4}b, respectively. Both quantities follow very similar scaling behavior, saying $Nu \sim Ra^\alpha$ with the exponents very close to each other. Therefore, the density flux ratio $\gamma=\left(\beta_T\langle v T\rangle\right)/\left(\beta_S\langle v S\rangle\right)$ is almost constant for different $Ra$, i.e. see Fig.~\ref{fig4}c. Constant $\gamma$ implies that the ratio of the density anomaly flux due to the heat flux to that due to the salinity flux exhibits very weak dependence on $Ra$. Moreover, $\gamma>1$ means that the net density anomaly flux is in the same direction as the scalar gradients along the top surface. 

In summary, we present a new type of thermohaline intrusion. Specifically, horizontal temperature and salinity gradients, which have compensated effects on density, can induce thermohaline intrusions in the water mass below, and interleaving layers with opposite moving directions emerge. Double diffusive mixing plays a crucial role in such process. Different types of DDC under the warm salty end and the cold fresh end produce downward fluxes of heat and salinity, which maintain the horizontal scalar differences at the lower part of the domain and the generation of new intrusion layers. The DDC mixing between different layers as they shift upward reinforces the interleaving as in the traditional DDC driven thermohaline intrusions. These intrusion currents provide an extra pass for horizontal heat and salinity transfer, which generates a density anomaly flux towards the warm and salty end. 

The thickness and the flow velocity of the interleaving layers exhibit similar behaviors as those within the water mass where temperature and salinity gradients exist. However, our results extend the circumstance where intrusions may occur, which is, they not only develop within the water mass of scalar gradients, but also in the fluid body at adjacent depth. Based on symmetry argument, one can expect that scalar gradients at bottom boundary should also induce similar intrusions above. Also, horizontal temperature and salinity gradients drive horizontal fluxes in the adjacent fluid layers, which may accelerate the decay of the horizontal scalar inhomogeneity. All these findings are of great interests and deserve experimental and observational verifications.

\

{\it Acknowledgements:} This work is supported by the Major Research Plan of National Natural Science Foundation of China for Turbulent Structures under the Grants 91852107 and 91752202. Y.Yang also acknowledges the partial support from the Strategic Priority Research Program of Chinese Academy of Sciences under the Grant No. XDB42000000.

\end{document}